# A Prediction Packetizing Scheme for Reducing Channel Traffic in Transaction-Level Hardware/Software Co-emulation

Jae-Gon Lee    Moo-Kyoung Chung    Ki-Yong Ahn    Sang-Heon Lee    Chong-Min Kyung
Department of EECS
Korea Advanced Institute of Science and Technology
Gusong-dong, Yusong-ku, Daejon, Korea
Tel: +82-42-862-6411
FAX: +82-42-862-6410
E-mail: {jglee, mystory, ahnky, shlee, kyung}@vslab.kaist.ac.kr

## Abstract

*This paper presents a scheme for efficient channel usage between simulator and accelerator where the accelerator models some RTL sub-blocks in the accelerator-based hardware/software co-simulation while the simulator runs transaction-level model of the remaining part of the whole chip being verified. With conventional simulation accelerator, evaluations of simulator and accelerator alternate at every valid simulation time, which results in poor simulation performance due to startup overhead of simulator-accelerator channel access. The startup overhead can be reduced by merging multiple transactions on the channel into a single burst traffic. We propose a predictive packetizing scheme for reducing channel traffic by merging as many transactions into a burst traffic as possible based on 'prediction and rollback.' Under ideal condition with 100% prediction accuracy, the proposed method shows a performance gain of 1500% compared to the conventional one.*

## 1. Introduction

### 1.1. Transaction-Level Modeling

Transaction-level modeling (TLM), usually described in SystemC, is a modeling style for SoC design with its focus on the external functional behavior of each block and inter-block communications without imposing excessive implementation details [7, 8]. According to the modeling of time, TLM is divided into two categories: architectural TLM and micro-architectural TLM [10, 11]. In architectural TLM, simulation time is only roughly modeled. This method is suitable for early stage prototyping of SoC. In micro-architectural TLM, the simulation time is modeled in a fully cycle-accurate manner. With micro-architectural TLM, we can verify SoC design in the early design stage with 100% cycle accuracy [3, 4]. This paper deals with simulation acceleration of micro-architectural TLM.

The simulation speed of TLM is much faster than that of RTL simulation. It is reported that the micro-architectural transaction-level models run at least two orders of magnitude faster than RTL models; simulation speeds of at least 100 kHz for a complete system simulation are readily achievable [2, 5, 10]. It is also possible to mix transaction level models with RTL models for gradual refinements but low simulation speed of RTL blocks limits the total simulation performance. In simulation accelerators, introduced to increase the RTL simulation speed, the limited throughput of the channel between simulator and simulation accelerator often restricts the overall performance gain.

### 1.2. Characteristics of the Simulator-Accelerator Channel

The channel between the simulator and the accelerator is composed of layers of API (Application Program Interfaces), device driver, and physical media each with static startup overhead. When PCI-based built-in simulation accelerator is used, experimental results show that startup overhead time is as big as 12.2 usec for each channel access whereas the payload times for simulator-to-accelerator and accelerator-to-simulator transfers are 49.95 nsec/word and 75.73 nsec/word each.[1]

To utilize the channel more efficiently, we need to send lots of data at a single time. But that's not possible with conventional simulation accelerators where the progress of simulator and accelerator are synchronized every simulation time. The amount of data to be sent between two simulation domains for a single simulation cycle is often too small to justify the startup overhead time of the channel; for an SoC design where each building block is interconnected via system bus, the amount of data does not exceed five words at a time. As a result, the transfer on the channel is composed of a series of short bidirectional transfers.

---

[1] Experimental results obtained with iPROVE[TM] tested with Pentium-4 2.8 GHz with 512 Mbytes of RAM and 32-bit PCI bus running at 33 MHz.

If we can remove one of the two types of transfers, we can merge remaining data transfers as a single transfer to minimize the number of channel accesses. This requires prediction of the states of the other simulation domain and recoveries from bad predictions. This paper deals with an optimistic simulator-accelerator channel usage based on 'prediction and rollback' to maximize the simulation speed when transaction-level models are executed by simulator while the behavior of RTL blocks is represented by accelerator.

## 2. Related Works

Parallel discrete event simulation (PDES), sometimes called distributed simulation, refers to the execution of a single discrete event simulation program on a parallel computer [6]. The concept of 'prediction and rollback' was first developed in PDES to extract maximum parallelism from a given problem. In the conservative method the progress of each process is synchronized at every simulation time. On the other hand, in optimistic method, each process proceeds assuming that there are no incoming messages. When a process receives a message with a time stamp smaller than the current simulation time, the process rollbacks to a previous state and sends negative messages to negate incorrect messages sent by the process.

The 'prediction and rollback' concept was first applied to SoC design by Yoo [12], where ARM prototype board models ARM processor and high level simulator models the behavior of hardware IPs. Yoo tried to minimize the channel accesses between the prototype board and simulator. But, his approach is based on high-level simulation without cycle-accurate behavior, which limits its practical use.

In [9], we proposed an "optimistic" simualtor-accelerator channel usage scheme based on 'prediction and rollback'. With the proposed method, one of the two *verification domain*s, i.e, *simulation domain* and *acceleration domain*, leads the other. The *leader* predicts the responses of the *lagger* so as to remove *lagger*-to-*leader* transfers on the channel. This allows to merge multiple *leader*-to-*lagger* transfers on the channel to minimize startup overhead of channel access. The proposed method has two operating modes, i.e., *Simulator Leading Accelerator* (*SLA*) and *Accelerator Leading Simulator* (*ALS*) depending on which of the two *verification do-*

*main*s leads the other. The duration of each *SLA* or *ALS* phase is named as a *transition* and a single *transition* is composed of four steps.

- *Run-Ahead step* (*RA step*) where *leader* proceeds predicting responses of *lagger*. Output of *leader* are not sent to *lagger* until *RA step* is over. Instead they are stored in *Leader Output Buffer* (*LOB*).
- *Follow-Up step* (*FU step*) where *lagger* follows up with the *leader*.
- Optional *RollBack step* (*RB step*) where *leader* rolls back to a previous state in case of prediction error detected.
- Optional *Roll-Forth step* (*RF step*) where *leader* runs again from the previous state to reach the progress of *lagger*.

Even though the proposed method can have great performance improvement compared to the conventional method, the paper focused on cases where two blocks residing in two different *verification domain*s communicate over static interconnections between them and failed to handle cases where multiple blocks are dynamically interconnected with a bus.

## 3. Problem Definition

We applied the "optimistic" channel usage pattern to SoC model where building blocks with different abstraction levels, i.e., TL and RTL, are interconnected with a system bus. It is assumed that the system bus follows the *Advanced High-performance Bus* (AHB) specification.[2] The bus dynamically utilizes interconnections among bus components, i.e., bus masters and bus slaves, to use the common resource for multidirectional data transfers. The dynamic utilization includes dynamic decision of data flow direction, dynamic decision of active bus components, dynamic decision of active interconnections, etc., which complicate the application of the "optimistic" channel usage pattern to SoC verification. The complications are summarized as follows.

1. How to split a single bus model into two sub-bus models without *combinatorial half loop*s between them?
2. How to limit number and types of signals between the two sub-bus models so that we can predict contents of at least one of the bidirectional data transfers between them?
3. Dynamic decisions on how to packetize bus signal values between the two sub-bus models.

---

2  AHB specification is one of the most popular and widely-used bus standards for embedded systems proposed by ARM [1].

4. Dynamic decisions among *SLA*, *ALS* and "conservative" operating modes.

We can meet the first constraint by letting each bus component to be present only in one of the two *verification domain*s, i.e., *simulation domain* and *acceleration domain*. As most bus specifications limit communication between bus components to take place only at edges of a clock signal, there can be no *combinatorial half loop* if each component is present only in a single *verification domain*.

We can meet the second constraint by limiting the subject of data transfers between simulator and accelerator only to elements of *minimal set of active bus signals*. *Set of bus signals* is a set of signals holding all the signals present in a bus specification. *Set of active bus signals* is a subset of *set of bus signals*, all of whose elements influences the operation of the bus. Signals driven by either by active bus master or active bus slave are elements of *set of active bus signals*.[3] Signals driven by arbiter and decoder are also elements of *set of active bus signals* as well as arbitration request signals driven by any of bus masters. *Minimal set of active bus signals* (*MSABS*) is a subset of *set of active bus signals*, values of whose elements can exclusively define the operations and states of the bus without redundant elements whose states can be deduced by combinations of states of the other elements of *MSABS*. Specifically, *MSABS* includes address (HADDR), control signals (HTRANS, HWRITE, HSIZE, HBURST, and HPROT), write data (HWDATA) of active bus master, read data (HRDATA), responses (HRESP, HREADY, and HSPLITx) of active bus slave, and arbitration request signals (HBUSREQx) of all bus masters.[4] Except for the arbitration request signals, all the elements of *MSABS* are related to data transfer on the bus and we call it *transaction bus signals*. In short, *MSABS* can be divided into two subsets of *set of transaction bus signals* and *set of arbitration request signals*. Among elements of *set of transaction bus signals*, values of address and control signals of active bus master can be deduced from their values at the start of a burst transfer on the target bus as their values either increase linearly over time or remain constant throughout a single burst transaction on the target bus. In other words, they are "predictable." Responses of active bus slave are also "predictable" as they just represent whether the active bus slave can handle bus transaction at a particular *target time*, which can be modeled with a simple producer-consumer model. This leaves read data and write data whose values cannot be effectively predicted, i.e., "non-predictable." As only one of the two data signals is active at a time, we set the source of the data flow as *leader* and the sink of it as *lagger* so that we do not need to predict values of data signals.

---

3 Active bus master refers to a bus master that is granted for bus access and active bus slave refers to a bus slave that is accessed by the active bus master.
4 Signal name in parentheses denotes the name of the corresponding signal under AHB specification [1]. It is assumed that arbitration priority and address maps of bus slaves are statically defined. This removes output signals of arbiter and decoder from the *minimal set of active bus signals* whose states (values) can be deduced from arbitration request signals and address signals.

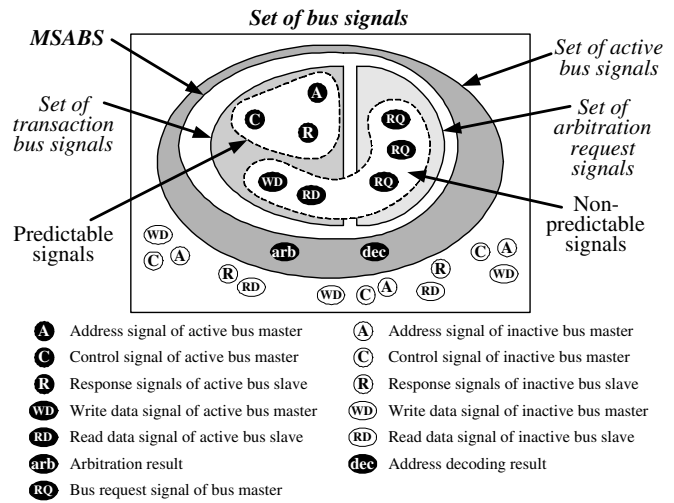

**Figure 1. Grouping of bus signals.**

Elements of *set of arbitration request signals* are "non-predictable" as we cannot predict whether a bus master will request for bus access at a particular *target time*. We cannot apply the same solution we used for data signals here as there can be sources of arbitration request signals, i.e., bus masters, on either side of simulator-accelerator channel. In other words, we should be able to predict the values of bus request signals driven by bus masters residing in *lagger*. But as the arbitration request signals contribute only to the generation of arbitration result signal, it suffices to be able to predict arbitration result signal value. In SoC designs where large amount of data flow in bursts between building blocks, the arbitration result tends to change only occasionally and we can effectively predict its value from its previous one. Figure 1 summarizes the above explanation. When there is any signal other than bus signals interconnecting two building blocks residing in different *verification domain*s, interrupt signal to be one of the most common examples, it should be treated the same as elements of *MSABS* and should be a subject of prediction, too.

The remaining problems related to dynamic use of simulator-accelerator channel are handled in Sect. 5.

## 4. Bus modeling

Figure 2 shows how a single SoC model is split into two *verification domain*s according to abstraction level of each component. For this purpose, a single bus model should be separately modeled by the two *verification domain*s and at the same time the two sub-bus models should be working closely together to model the behavior of a single target bus model. This is achieved by two *half bus model*s (*HBM*s) and two *Channel Wrapper*s (*CW*s) in Fig 2. *HBM*s residing in *simulation domain* and *acceleration domain* are called *Half Bus Model for Simulator* (*HBMS*) and *Half Bus Model for Accelerator* (*HBMA*), respectively. *CW*s residing in *simulation domain* and *acceleration domain* are called *Channel Wrapper for Simulator* (*CWS*) and *Channel Wrapper for Accelerator* (*CWA*), respectively. The structure of each *HBM* is just the same as conventional bus models. It holds an arbiter and a decoder and is connected to bus masters and slaves either in transaction-level or in pin-level. In either *verification domain*, *CW* mimics the behavior of bus masters and slaves present in the other *verification domain*. This allows each *HBM* in either *verification domain*s to function just the same as the target bus model does. The two *CW*s exchange active bus signal values over the simulator-accelerator channel. When one *CW* can predict all the signal values to read from the other *CW*, it replaces the read access with prediction for "optimistic" operation.

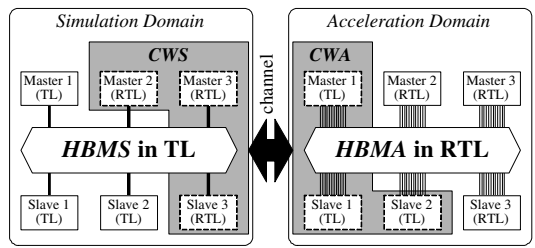

**Figure 2. Bus modeling.**

## 5. Operations of Channel Wrapper

The operation of *CW* is realized with a state diagram shown in Fig. 3. Each of the two *CW*s holds its own state and their combination represents each step of a *transition*. Operations of *CW* can be grouped into six paths denoted as *F* (roll-Forth), *P* (Prediction), *S* (Synchronization), *L* (Lagger), *R* (Report), and *C* (Conservative) paths in Fig. 3. Each path represents operations of a *CW* for a single simulation cycle. Except for the case when operation of a *CW* is blocked at some blocking read operations, state of the *CW* flows from *START* to *END* at every positive edge of a clock signal. This is called a *unit cycle operation* of *CW*. Let's assume that two *CW*s start running in conventional operating modes: "conservative" cycle-by-cycle synchronization. During this time, two *CW*s take *C-path* together leapfrogging each other. At some moment, one of the two *CW* realizes that it can predict the response of the other and takes *P-path* instead of *C-path*. This is the start of a new *transition*. Even though *leader* takes *P-path*, it does not run in "optimistic" way (no predictions are made) when this is the first time to take *P-path* for a *transition*. This is to store the state of *leader* before taking "optimistic" operations for possible *rollback*s in the future. Leading *CW* takes *rb_store* state (denoted as *P-5* in Fig. 3) to register a state store after current *unit cycle operation* is over.[5] After that *leader* takes *C-path* for "conservative" operation (*P-6*). The "optimistic" channel usage starts when *leader* comes back to *P-path* again at the very next *unit cycle operation*.[6] Now *leader* takes different path in the *P-path* to write output values of *leader* to *LOB* and to predicts the response of *lagger*. The prediction result is stored in the *LOB* along with output data of *leader*. The prediction results is reflected on the current *verification domain* before current *unit cycle operation* is over; predicted signal value is sent to the current *verification domain* as if it was read from the other *verification domain*. This continues until *leader* cannot predict the response of *lagger*. When *leader* cannot proceed without synchronization, *leader* takes *S-path* instead of *P-path*. In *S-path*, the contents of *LOB* are flushed to *lagger* (*S-2*) and *leader* waits for the reply from *lagger* in *Get response* state (*S-3*).

Until then, *lagger* waits for *leader* to write data in *Read input data* state (*C-3*). Now that *LOB* is flushed, *lagger* can get out of the blocking read operation to finish the simulation cycle and to take *L-path* at the next simulation cycle. *Lagger* checks a single prediction every time it reaches *Prediction check* state (*L-1*). If the prediction coincides with the actual response of *lagger*, *lagger* reads another leader-to-lagger data and finishes the *unit cycle operation*. When all *leader*-to-lagger data are consumed, i.e., all the predictions are correct, *lagger* takes *R-path*, and reports this to *leader*, which has been waiting for this response in *S-3*.[7] In *R-path*, the output of *lagger* is sent to *lagger* immediately (*R-2*) and *lagger* once again waits for the leader-to-lagger at *Read input data state* (*R-3*). And this is the end of a successful *transition*.

When a prediction failure is detected in *Prediction check* state in *L-path*, it is reported to *leader* immediately (*L-5*). After reporting the prediction failure, *lagger* waits for the leader at *L-6*. Upon receiving this message, *leader* takes prediction failure path in *S-path*. First, *leader* stores the last response of *lagger* (*S-5*), which *leader* failed to predict, and requests state restore (*S-6*). After the *unit cycle operation* of leading *CW* is over, the state of *leader* is rolled back to a state *P-path* stored in the past.[8] Now *leader* takes *F-path* for roll-forth operations. The operations of *F-path* resemble those of *P-path* except that *F-path* does not write output signal values to *LOB*. After iterating taking *F-path* for the number of successful predictions, the *transition* is over.

---

5  For simulator, the state is stored after all operations for the current *simulation time* is over and all the variables are stabilized. This is to save memory requirements for the state storage. For accelerator, the state is stored as soon as ACW reaches *P-5* as the signal values of accelerator is stabilized as soon as clock signal toggles [9].

6  If *leader* cannot predict the response of *lagger* at the next simulation cycle, the *transition* is over and there is no optimistic channel usage and the proposed method works just the same as conventional method with unnecessary state store overhead spent at the previous *unit cycle operation*.

7  *Lagger* can figure out that the last leader-to-lagger data is reached as the last leader-to-lagger data does not contain prediction. The last *unit cycle operation* of leading *CW* does not predict the state of *lagger* as it tries to read it from *lagger* as conventional method does.

| Roll step | Leader state | Lagger state | Description |
|---|---|---|---|
| RA step | P path | L,R,C path | Leader predicts responses of lagger to remove read transaction except for the first time in P-path. |
| FU step | S path | L path | Lagger follows up *leader* until either they are synchronized or prediction error is found. |
| | | R | Lagger reports that all the predictions were correct. |
| RB step | S path | L path | States of *leader* gets rolled back to a prev. state |
| RF step | F path | L path | Leader follows up the progress of *lagger*, which is waiting for *leader*. |

**Table 1. Roll steps and CW states.**

Operations of each path and their relations to *transition* steps are summarized in Tbl. 1.

## 6. Experimental Results

The performance of the proposed idea is highly sensitive to the prediction accuracy. Low prediction accuracy degrades performance by increasing not only the number of rollbacks but also the number of state restores and state stores. But the biggest degradation comes from the increased number of clock cycles to be processed by *leader* and channel accesses as shown in Tbl. 2. Table 2 shows the effect of prediction accuracy to the duration of time spent by each component operation under *ALS* operating mode. We assumed simulator speed of 1,000 kcycles/sec, accelerator speed of 10 Mcycles/sec, *LOB* depth of 64 and 1,000 rollback variables. $T_{sim.}$ and $T_{acc.}$ stand for the average time spent by simulator and accelerator to model the behavior of a target SoC model for a single target clock cycle, respectively. $T_{store}$ and $T_{restore}$ stands for the time spent in stor-ing and restoring the state of *leader*. $T_{store}$ stands for the time spent in accessing simulator-accelerator channel. Conventional method has a simulation speed of 38.9 kcycles/sec under the same environment. The proposed method has performance gain of 16.75 when all the predictions are correct. The performance of ALS drops as the prediction accuracy drops. When it equals to 10%, the performance of ALS is about the same as that of conventional method.

---
[8] The state of *CW* is not rolled back.

Figure 4 shows the performance estimation of *ALS* under four different configurations with two different simulator speed and two *LOB* depths. As the performance of hardware-based simulation accelerator is independent of design sizes, we kept accelerator speed to be constant. The bigger the simulator performance gets, we get the more performance gain from the proposed method. *LOB* depth decides the maximum number of predictions and tends to accelerate the performance gain of the proposed idea when the prediction accuracy is high. On the other hand, it degrades the performance gain when the prediction accuracy is low.

The performance of *SLA* has similar tendencies: maximum performance gain of 3.25 and 15.34 for simulation performance of 100 kcycles/sec and 1,000 kcycles/sec, each. But it was found that *SLA* suffers more from low prediction accuracies. This is because relatively low operating speed of *simulation domain* compared to that of *acceleration domain* enlarges the effect of the most dominant factor of performance degradation, i.e., time spent by *leader*. *SLA* has the same simulation performance as the conventional method when the prediction accuracy is 98%[70%] assuming that the simulation performance is 100 kcycles/sec[1,000 kcycles/sec].

| Prob. | 1.000 | 0.990 | 0.960 | 0.900 | 0.800 | 0.600 | 0.300 | 0.100 |
|---|---|---|---|---|---|---|---|---|
| $T_{sim.}$ | 1.0e-6 | 1.0e-6 | 1.0e-6 | 1.0e-6 | 1.0e-6 | 1.0e-6 | 1.0e-6 | 1.0e-6 |
| $T_{acc.}$ | 1.0e-7 | 1.6e-7 | 2.9e-7 | 4.9e-7 | 8.1e-7 | 1.5e-6 | 2.4e-6 | 3.0e-6 |
| $T_{store}$ | 4.69-10 | 7.6e-10 | 1.6e-9 | 3.3e-9 | 6.2e-9 | 1.2e-8 | 2.1e-8 | 2.7e-8 |
| $T_{rest.}$ | 0 | 2.9e-10 | 1.2e-9 | 2.9e-9 | 5.7e-9 | 1.2e-8 | 2.0e-8 | 2.6e-8 |
| $T_{ch.}$ | 4.3e-7 | 6.8e-7 | 1.5e-6 | 2.9e-6 | 5.4e-6 | 1.1e-5 | 1.8e-5 | 2.3e-5 |
| Perform. | 652k | 543k | 363k | 226k | 138k | 76.7k | 46.1k | 36.7k |
| Ratio | 16.75 | 13.97 | 9.33 | 5.80 | 3.56 | 1.91 | 1.19 | 0.94 |

**Table 2. Performance of ALS.**

## 7. Conclusion

Micro-architectural transaction-level modeling enabled early stage SoC verification with full cycle accuracy and fast simulation speed. Usually, transaction-level models are gradually refined to RTL models. But as the simulation speed of RTL models are too slow, the verification speed degrades as the proportion of RTL blocks increases. We can alleviate this problem with simulation accelerator, but now the throughput of simulator-accelerator channel limits the simulation performance. This gets worse as transactions on the channel are composed of series of short transfers, which suffers from static startup overhead of the channel. To minimize the effect of startup overhead and to get maximum simulation speed, we introduced the concept of 'prediction and rollback' to the synchronization between simulator and accelerator. With this method, the progress of simulator and accelerator are not synchronized at every simulaton time, but they are synchronized only when it is inevitable for cycle accurate behavior. We adopted the concept to a system bus model to get a profound performance gain when prediction accuracy is high.

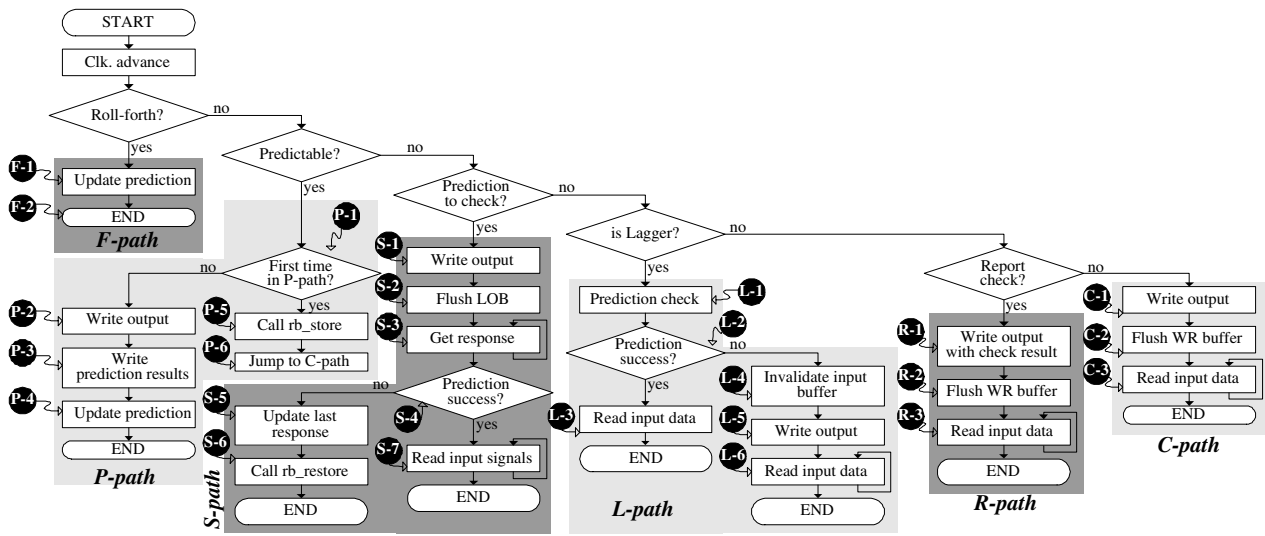

**Figure 3. Operations of Channel Wrapper. Each path of which is assigned to an alphabetical letter.**

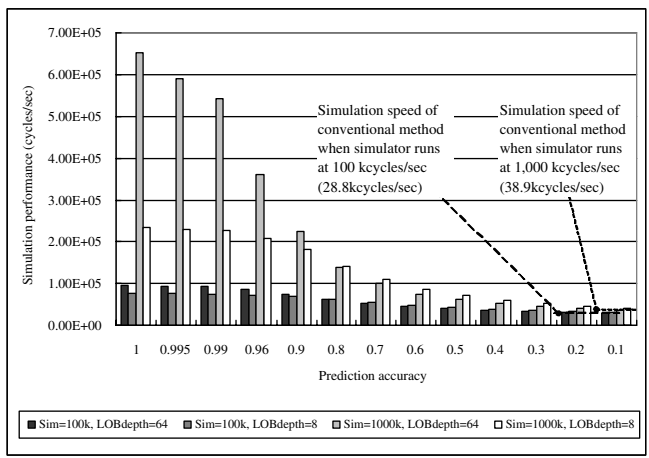

**Figure 4. Experimental results.**